\documentstyle{mn}
\input psfig.tex
\newcommand{\ltsima} {$\; \buildrel < \over \sim \;$}
\newcommand{\gtsima} {$\; \buildrel > \over \sim \;$}
\newcommand{\lta} {\lower.5ex\hbox{\ltsima}}
\newcommand{\gta} {\lower.5ex\hbox{\gtsima}}

\def\refitem{\par\parskip 0pt\noindent\hangindent 20pt}

\title[Rapid variability of blazars]
{Rapid variability in the synchrotron self Compton model for blazars}

\author[M. Chiaberge and G. Ghisellini]
{Marco Chiaberge$^{1,2}$ and Gabriele Ghisellini$^3$ \\
$^1$ Osservatorio Astronomico di Torino, Strada Osservatorio, 20,
I-10025 Pino Torinese,  Italy \\
$^2$ S.I.S.S.A., Scuola Internazionale Superiore di Studi Avanzati,
via Beirut 4, I-34014 Trieste, Italy \\
$^3$ Osservatorio Astronomico di Brera, V. Bianchi, 46, I-22055 Merate, Italy}

\date{Received ***; in original form ***}

\begin{document}

\maketitle

\begin{abstract}
Blazars are characterized by large amplitude and fast variability,
indicating that the electron distribution is rapidly changing,
often on time scales shorter than the light crossing time.
The emitting region is sufficiently compact to let radiative 
losses dominate the cooling of high energy electrons.
We study the time dependent behaviour of the electron distribution
after episodic electron injection phases, and calculate the observed
synchrotron and self Compton radiation spectra.
Since photons produced in different part of the source have different
travel times, the observed spectrum is produced by the electron distribution
at different stages of evolution.
Even a homogeneous source then resembles an inhomogeneous one.
Time delays between the light curves of fluxes at different frequencies
are possible, as illustrated for the specific case of the BL Lac object 
Mkn 421.

\end{abstract}

\begin{keywords}
Galaxies: active --- BL Lacertae objects: individual: Mkn 421 ---
radiation mechanisms: nonthermal --- methods: numerical --- X--rays: galaxies
\end{keywords}

\section{Introduction}

Variability is one of the defining properties of blazars, characterized  by 
variations of their flux even of two orders of magnitude in time scales of 
years, and smaller changes, but still up to factor 2, in hours/days time
scales.
Observational efforts to characterize the variability behaviour, which started
soon after the discovery of blazars, have been intensified in recent years, 
to study very fast fluctuations and possible correlation of fluxes at 
different frequencies (for a review see Ulrich, Urry \& Maraschi, 1997).
This will hopefully shed light on the location of the emitting regions 
and on the nature of the acceleration mechanisms.
Theoretical efforts, however, lag behind observations:
the  classical paper on time dependent synchrotron and Compton flux is 
still Kardashev (1962), which solves the continuity equations
for the particle distribution in cases where the cooling time $t_{cool}$
can be considered longer than the light crossing time $R/c$ for any energy.
Recently, Mastichiadis \& Kirk (1997) and Atoyan \& Aharonian (1997),
apply the results of solving the continuity equation for the electron
distribution to the variations seen in the BL Lac object Mkn 421 and 
the galactic superluminal source GRS 1915+105.
In both cases, only the variability on time scales longer than $R/c$
were studied.
On the other hand, significant variability often happens on time scales 
extremely short in highly luminous (and compact) objects.
This implies that the cooling time for the highest energy electron
may well be shorter than $R/c$, even once the effect of Doppler
boosting and blueshift is accounted for.
This is also indicated by the behaviour of the light curves as seen, e.g., 
in the X--rays and in the optical, showing a quasi--symmetric behaviour, 
with rise and decay time scales approximately equal
(see i.e. Urry et al. 1997, Ghisellini et al. 1997, Massaro et al. 1996,
Giommi et al. 1998), indicating that both times are connected to
the light travel time across the source $R/c$, and therefore suggesting that
the cooling times of the emitting electrons are shorter.
This in turn implies that the electron distribution, at least at these
energies, is significantly changing on time scales shorter than $R/c$.

In these cases the knowledge of the time evolution of the particle
distribution of the emitting electrons is not enough to construct
theoretical light curves for the $observed$ fluxes, since light travel time
effects play a crucial role.
The observer in fact sees, at any given time, photons produced in different
part of the source, characterized by a particle distribution of a different age.
The observed flux is then the sum of the emission produced by
{\it different} particle distributions.

In \S 2 of this paper we present our method of solving the continuity
equation for the emitting particle distribution, together with the
assumptions made.
In \S 3 we discuss illustrative examples of the time behaviour of
the particle and photons distributions, having care to evidentiate
the effects introduced by the different photon travel times.
Besides considering simple illustrative cases where the injection of
`fresh' relativistic particles is assumed to occur simultaneously throughout 
the source, we also study more realistic cases which simulate the 
injection occuring when a shock front travels down a region of a jet.
In \S 4 we apply our model to Mkn 421, to show how it is possible, 
even in our simplified model, to explain
the time lags observed in the hard and soft X--ray light curves.
Finally, in \S 5 we discuss our findings.

\section{The model}
\subsection{Assumptions}

We assume that the emission is produced by a distribution
of relativistic electrons injected in a region of typical dimension
$R$ embedded in a tangled magnetic field $B$, at a rate $Q(\gamma)$ 
[cm$^{-3}$ s$^{-1}$] ($\gamma$ is the Lorentz factor). 
Electrons lose energy by emitting synchrotron and synchrotron self-Compton
radiation (SSC); they can also escape from the emitting region in a
time scale $t_{esc}$, assumed to be independent of energy. 

Our main purpose is to apply our calculations to the study of the blazars
short time scales variability, especially in frequency bands around and 
above the peaks of blazars spectral energy distribution.
In these spectral regions synchrotron self-absorption is negligible,
and consequently we neglect this heating effect 
in the kinetic equation (e.g. Ghisellini, Guilbert \& Svensson, 1988).
Photon--photon collisions, producing electron--positron pairs, are
also neglected, since we will deal with sources of small compactness
$\ell\equiv L\sigma_T/(Rmc^3)$ (Maraschi, Ghisellini \& Celotti, 1992)

The continuity equation governing the temporal evolution of the 
electrons distribution $N(\gamma,t)$ [cm$^{-3}$] is 
\begin{equation}
\frac{\partial N(\gamma,t)}{\partial t} = \frac{\partial}{\partial\gamma} 
\left[ \dot\gamma(\gamma,t) N(\gamma,t)\right] + Q(\gamma,t) - 
\frac{N(\gamma,t)}{t_{esc}}
\label{cont}
\end{equation}
where $\dot\gamma= \dot\gamma_{S} + \dot\gamma_{C}$ is the total cooling rate,
given by
\begin{equation}
\dot\gamma = \frac{4}{3} \frac{\sigma_T c}{m_e c^2} [ U_B+U_{rad}(\gamma,t)] 
\gamma^2
\end{equation}
where $\sigma_T$ is the Thomson cross section, $U_B$ is the magnetic field
energy density and $U_{rad}(\gamma,t)$ is the energy density of the radiation
field. 

The cooling time scale is assumed to correspond to synchrotron and
self-Compton radiative losses:
\begin{equation}
t_{cool} = { 3 m_ec^2 \over 4 \sigma_T c \gamma [U_B+U_{rad}]} 
\end{equation}
This relation can be rewritten by assuming that all radiation
energy density is available for scattering, and that it
corresponds to the compactness $\ell=4\pi R \sigma_T U_{rad}/ (m_ec^2)$:
\begin{equation}
{t_{cool} \over R/c} = { 3 \pi \over \gamma \ell [1 + U_B/U_{rad}]} 
\end{equation}
Note that for compactness values in the range $10^{-4}$--$10^{-1}$,
typical for the IR to $\gamma$--ray emission of blazars,
the high energy electrons cool faster than $R/c$.

\subsection{Numerical method}

We numerically solve equation (\ref{cont}) adopting the fully implicit
difference scheme proposed by Chang \& Cooper (1970), modified for 
our purposes, since we are dealing with a continuity equation with 
injection and escape terms and no heating. 
The Chang \& Cooper (1970) scheme allows to find more stable, 
non-negative and particle number conserving solutions.
Moreover that scheme significantly 
reduces the number of meshpoints required to obtain accurate solutions, and
it is also appropriate for including heating terms (Ghisellini, Guilbert \&
Svensson, 1988).

For all our runs we use an energy grid with equal logaritmic resolution:
the energy meshpoints are defined as 
\begin{equation}
\gamma_j = \gamma_{min}
\left( \frac{\gamma_{max}}{\gamma_{min}} \right)^\frac{j-1}{j_{max}-1}
\end{equation}
where $j_{max}$ is the meshpoints number, and the energy intervals are
$\Delta \gamma_j = \gamma_{j+1/2} -\gamma_{j-1/2}$. Quantities with the
subscript $j\pm 1/2$ are calculated at half grid points.
In the performed simulations a grid of 200 points has been used
both for particle energy and photon frequency.

In order to discretize the continuity equation we define 
\begin{equation}
N^i_j = N(\gamma_j, i\Delta t)
\end{equation}

\begin{equation}
F^{i+1}_{j\pm 1/2} = \dot\gamma^i_{j \pm 1/2} N^{i+1}_{j\pm 1/2}
\end{equation}
and equation (\ref{cont}) can be written as
\begin{equation}
\frac{N^{i+1}_j - N^i_j}{\Delta t} = \frac{F^{i+1}_{j+1/2} - 
                                     F^{i+1}_{j-1/2}}{\Delta \gamma} + Q^i_j - 
                                     \frac{N^{i+1}_j}{t_{esc}}
\end{equation}
In this specific case we have $N_{j+1/2}\equiv N_{j+1}$ and $N_{j-1/2}
\equiv N_{j}$,
according to the prescriptions of Chang \& Cooper (1970).
We can now rewrite the continuity equation as
\begin{equation}
V3_j N^{i+1}_{j+1} + V2_j N^{i+1}_j + V1_j N^{i+1}_{j-1} = S^i_j
\label{sys}
\end{equation}
where the V coefficients are
\begin{eqnarray}
V1_j & = & 0 \nonumber \\
V2_j & = & 1+ \frac{\Delta t}{t_{esc}} + \frac{\Delta t \, \dot\gamma_{j-1/2}}
{\Delta \gamma_j} \\
V3_j & = & -\frac{\Delta t \, \dot\gamma_{j+1/2}}{\Delta \gamma_j} \nonumber
\end{eqnarray}
and
\begin{equation}
S^i_j = N^i_j + Q^i_j \, \Delta t \, .
\end{equation} 
The system of equations (\ref{sys}) forms a tridiagonal matrix,
and it is solved numerically (e.g. Press et al. 1989). 
We tested our method with the analitic 
solutions given by Kardashev (1962) in the case of synchrotron radiation only
and injection of a constant power--law and monoenergetic distributions.

We then calculate the synchrotron emissivity $\epsilon_s(\nu,t)$ 
[erg s$^{-1}$ cm$^{-3}$ sterad$^{-1}$ Hz$^{-1}$] of each distribution 
$N(\gamma_j,t_i)$ with 
\begin{equation}
\epsilon_s(\nu,t) = \frac{1}{4\pi} \int^{\gamma_{max}}_{\gamma_{min}} d\gamma
\, N(\gamma,t) \, P_s(\nu,\gamma)
\end{equation}
In the above formula $P_s(\nu,\gamma)$ 
[erg s$^{-1}$ Hz$^{-1}$ sterad$^{-1}$] is the single particle synchrotron 
emissivity averaged over an isotropic distribution of pitch angles 
\begin{eqnarray}
P_{s}(\nu,\gamma) & = & \frac{3 \sqrt{3}}{\pi} \frac{\sigma_{T} c 
U_{B}}{\nu_{B}} t^{2} \times \\ 
&  \times & \left\{K_{4/3}(t) \, K_{1/3}(t)- \frac{3}{5} t \, 
 [K_{4/3}^{2}(t)-K_{1/3}^{2}(t)] \right\} \nonumber
\end{eqnarray}
(Crusius \& Schlickeiser 1986; Ghisellini, Guilbert \& Svensson 1988), 
where $t=\nu/(3\gamma^2\nu_B)$, $\nu_B = e B/(2\pi m_e c)$, $K_a(t)$ 
is the modified Bessel function of order $a$.
We then calculate the synchrotron radiation field $I_s(\nu ,t)$ 
[erg s$^{-1}$ cm$^{-2}$ sterad$^{-1}$ Hz$^{-1}$] using the transfer equation
\begin{equation}
I_s(\nu,t)=\frac{\epsilon_s(\nu,t)}{k(\nu,t)} \left[ 1- e^{-k(\nu,t) R} \right]
\end{equation}
here $k(\nu,t)$ [cm$^{-1}$] is the absorption coefficient 
(e.g. Ghisellini \& Svensson, 1991)
\begin{equation}
k(\nu,t) = - \frac{1}{8 \pi m_e \nu^2} \int^{\gamma_{max}}_{\gamma_{min}}
              \frac{N(\gamma,t)}{\gamma p}
              \frac{d}{d\gamma}[\gamma p \, P(\gamma,\nu)] 
\end{equation}
where $p=(\gamma^2-1)^{1/2}$ is the particle momentum in units of $m_e c$.
In order to calculate the inverse Compton radiation field,
we make the assumption that the synchrotron radiation
instantaneously fills the whole of the emitting region (see also \S 
\ref{slices}) and we take into account 
the Klein-Nishina decline using the following approximation (Zdziarsky, 1986)
\begin{equation}
\sigma = \left\{ \begin{array}{ll}
                 \sigma_{T} & \mbox{for $\gamma x<3/4$}\\
                 0  & \mbox{for $\gamma x> 3/4$}
                 \end{array}
         \right. 
\label{knappr}
\end{equation} 
where $x = h\nu/m_e c^2$. 
We approximate the energy density of the synchrotron radiation 
$U_{rad,syn}(\gamma,t)$ by setting
\begin{equation}
U_{rad,syn}(\gamma,t)= \frac{4 \pi}{c}
                   \int_{\nu_{s,min}}^{\nu_{max}(\gamma)} d\nu \, I_{s}(\nu,t) 
\end{equation}
and the integration limits are defined as $\nu_{min}=\nu_{s,min}$ (minumum
synchrotron emitted frequency)
and $\nu_{max}(\gamma) = \min \left[ \nu_{s,max}, 3 m_e c^2/(4 h \gamma)
\right]$.\\
The inverse Compton emissivity can be calculated, with the above assumptions,
following Rybicki \& Lightman (1979)
\begin{equation}
\epsilon_{c}(\nu_{1}) =  
         \frac{\sigma_{T}}{4}  \int^{\nu_{0}^{max}}_{\nu_{0}^{min}}
                     \! \frac{d\nu_{0}}{\nu_{0}}
                 \int^{\gamma_2}_{\gamma_1} \! 
                 \frac{d\gamma}{\gamma^{2}\beta^{2}}
             N(\gamma) \, f(\nu_{0},\nu_{1}) \, \frac{\nu_{1}}{\nu_{0}}
                  \, I_{s}(\nu_{0}) 
\end{equation}
where $\nu_0$ is the frequency of the incident photons, $\nu_1$ is the
frequency of the scattered photons, $\beta=v/c$, and the integration limits are
\begin{equation}
\gamma_1= \max \left[ \left( \frac{\nu_{1}}{4\nu_{0}} \right)^\frac{1}{2} \, ,
           \gamma_{min} \right] 
\end{equation}
\begin{equation}
\gamma_2= \min \left[  \gamma_{max} \, , 
           \frac{3}{4} \frac{m_{e} c^{2}}{h \nu_{0}} \right] \, \\
\end{equation}
and where $\nu_0^{max}$ and $\nu_0^{min}$ are the extreme frequencies
of the synchrotron spectrum.
The function $f(\nu_{0},\nu_{1})$ is the spectrum produced by the single 
electron, scattering monochromatic photons of frequency $\nu_0$ 
(see e.g. Rybicki \& Lightman, 1979)
\begin{equation}
f(\nu_{0},\nu_{1})= \left\{ \begin{array}{ll} 
            (1+\beta)\frac{\nu_{1}}{\nu_{0}} - (1-\beta);
            & \mbox{$\frac{1-\beta}{1+\beta}\leq \frac{\nu_{1}}
             {\nu_{0}}\leq 1$}\\
            (1+\beta) - \frac{\nu_{1}}{\nu_{0}} (1-\beta);
            & \mbox{$1\leq \frac{\nu_{1}}{\nu_{0}}\leq \frac{1+
            \beta}{1-\beta}$}\\  
            0 & \mbox{otherwise}
                     \end{array}
              \right.
\end{equation}

The inverse Compton radiation field is simply obtained
as $I_c(\nu_1)= \epsilon_c(\nu_1) \, R$.\\

\subsection{Slices}
\label{slices}

The knowledge of $N(\gamma, t)$ is not sufficient to properly reproduce 
the variability behaviour of blazars: observations in the optical
band and at higher frequencies strongly suggest that the injection and/or
the cooling times can be shorter than the light crossing time $R/c$. 
In this case the particle distribution evolves more rapidly than
$R/c$ {\it and the observer will see a convolution of different
spectra, each produced in a different part of the source}. 
Initially the observer only sees the emission coming from fresh electrons
located in the region (`slice') closest to her/him; then the innner parts
of the source become visible, also showing `young' spectra, while
electrons in the front slices are evolving.
After a time $R/c$ all the emitting region will be visible:
the back of it with fresh electrons and the front of it with older electrons.

In order to reproduce this effect we can divide the source of size $R$ into 
$n$ slices of equal thickness $R_{sl}$ and then sum the contribution 
of each slice to obtain the correct observed spectrum.
Note that the number of slices has to be chosen in order to have 
$R_{sl}/c < t_{min}$, where $t_{min}$
is the shorter among the relevant time scales (cooling, injection and 
escape time):
in this way each slice can be considered an {\it homogeneous} region, while 
the different slices are observed in a different evolution time, 
resembling an {\it inhomogeneous} emitting region.

We consider the case in which all slices have equal volumes,
and assume a `cubic' geometry, with the line of sight placed at 90$^\circ$
with respect to one face of the cube.
This angle, besides corresponding to the simplest case, 
is transformed, in the lab frame, into a viewing angle of $\sim 1/\Gamma$,
if the source moves with a Lorentz factor $\Gamma$, which is
appropriate for blazars.
Extension to different geometries (i.e. cylinder, sphere, etc)
is trivial, by properly weighting each slice volume.
We will see that even in the simplest, `cubic', case, time-lags among 
light curves at different frequencies are clearly observable.

We note that the light crossing time effects should be taken into 
account also for the correct calculation of the inverse Compton radiation:
synchrotron photons in fact require a time $R/c$ to fill the 
whole of the emitting region.
This effect, relatively important for transient high energy phenomena in 
Compton dominated sources, is however negligible for 
magnetic dominated objects, and we neglect it in the present 
paper, due to the computing time it needs to be properly calculated.
We plan to improve our numerical code in the future.
We expect that this effect will cause a time delay between synchrotron 
and self-Compton fluxes.


\section{Simulations}

\subsection{Gaussian injection with $t_{inj}\ll R/c$}
\label{gauss}

We first examine the case of a narrow Gaussian
electron distribution continuously injected for a time $t_{inj} \ll R/c$
in a region of size $R=10^{16}$ cm embedded in a tangled magnetic field
$B=1$ Gauss. 
The injected distribution peaks at $\gamma=10^5$. 
We measure the total injected power with the corresponding 
compactness, defined as $\ell_{inj}\equiv L_{inj} \sigma_{T}/(R m_e c^3)$: 
in this case we use $\ell_{inj}=10^{-3}$.
The injection stops at $t=R/(10 \, c)$, and we set the escape time 
$t_{esc}=1.5 R/c$. This is the value of $t_{esc}$ used in all simulations, 
unless otherwise noted.

This case could represent a physical situation of a narrow region of the jet
invested by a perturbation which activates the region. In this case $R$ 
corresponds to the dimension perpendicular to the jet axis, while the width
of the region could correspond to $\sim t_{inj}c \ll R$. Besides possibly
representing a real situation (even if highly idealized), this case is
useful to clearly show the light crossing time effects.
A more detailed discussion of the simulation of a region powered by a shock
will be presented in section \ref{shock}.

In order to better understand the different behaviours of the evolution 
of both particle and photons distributions, we can roughly divide the electron 
distribution and the emitted synchrotron spectrum into three main energy ranges
depending on the characteristic involved time scales, with respect to $R/c$:
\begin{enumerate}
\item low energy 
($\gamma \lta 2\times 10^2$, $\nu_{syn} \lta 10^{11}$ Hz), in which
the cooling time is at least one order of magnitude longer than $R/c$, 
so particle escape is the leading effect;
\item medium energy ($2\times 10^2 \lta \gamma \lta 2\times 10^4$, \linebreak
$10^{11}{\rm Hz} \lta \nu_{syn} \lta 10^{15}$ Hz), in which the electron 
cooling time is comparable to $R/c$, 
so that the cooling time is comparable to the escape time;
\item high energy 
($\gamma \gta 2\times 10^4$, $\nu_{syn} \gta 10^{15}$Hz)
in which $t_{cool} \ll R/c$ (by at least one order of magnitude), 
so that the particle distribution evolves more rapidly than $R/c$.
\end{enumerate}

\begin{figure}
\psfig{file=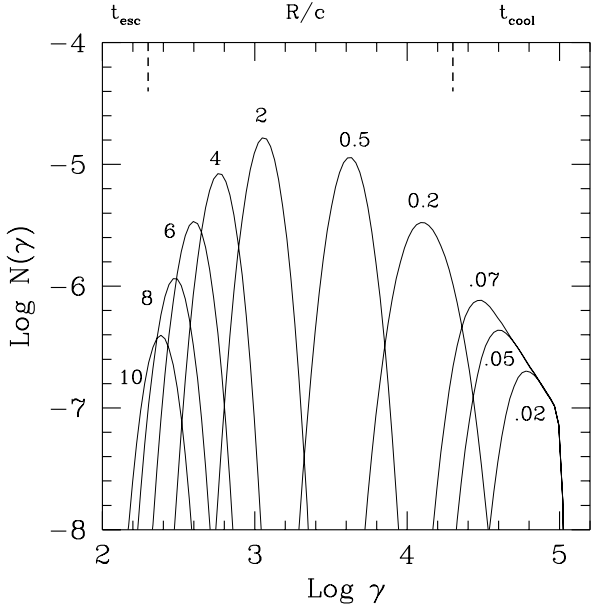,width=8.5truecm,height=8.5truecm}
\caption[h]{Evolution of the particle distribution $N(\gamma)$
corresponding to an injection of particles distributed in energy
as a Gaussian, centered at $\gamma=10^5$, for $t_{inj}=0.1 R/c$.
Labels indicate time after the beginning of the injection, 
in units of $R/c$ (in the comoving frame).
The escape time scale is $t_{esc}=1.5 R/c$.
The energy range has been divided according to the relevant 
time scales (top labels, see text).}
\label{fig1}
\end{figure}

\begin{figure}
\psfig{file=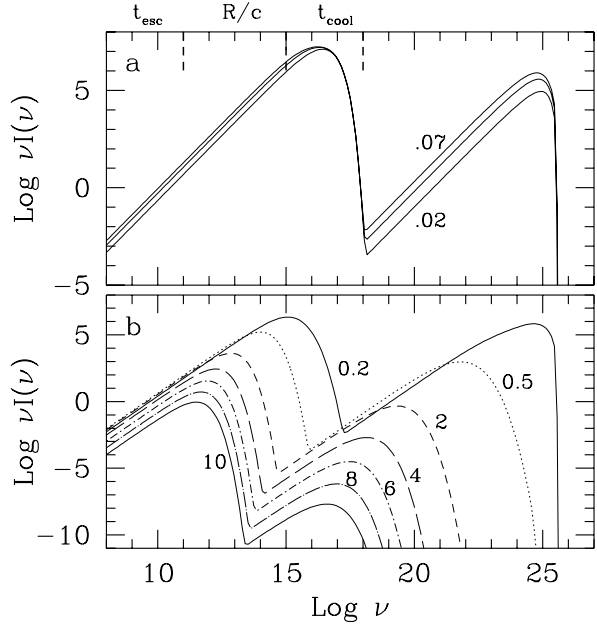,width=8.5truecm,height=8.5truecm}
\caption[h]{Evolution of the synchrotron self-Compton spectrum emitted
by the particle distributions shown in Fig. 1, during a) the injection phase,
and b) for $t>t_{inj}$. Labels indicate the time after the beginning
of the injection, in units of $R/c$ (in the comoving frame).
These spectra {\it do not} take into account light crossing time effects.}
\label{fig2}
\end{figure}

\begin{figure}
\psfig{file=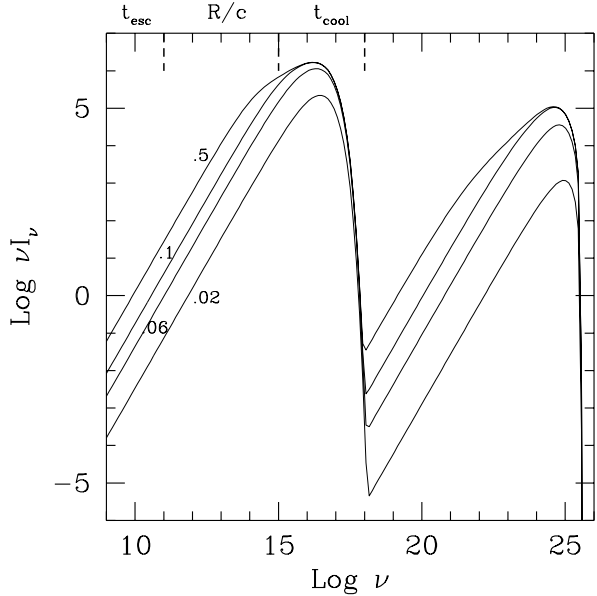,width=8.5truecm,height=8.5truecm}
\caption[h]{Evolution of the synchrotron self-Compton spectrum emitted
by the particle distributions shown in Fig. 1, {\it taking into account}
light crossing time effects. Note that the intensity is still rising at 
$t=0.5 \, R/c$ for frequencies under both the synchrotron and the 
Compton peaks, although the injection stops at $t=0.1 \, R/c$.}
\label{fig3}
\end{figure}

In Fig. \ref{fig1} we plot the electron distributions $N(\gamma)$ at different
times, with high temporal resolution for the first instants, and lower 
temporal resolution for the rest of the evolution, when the cooling effects 
are much slower.

The higher energy part of the
distribution reaches the equilibrium state in a very short time, and
does not vary during the injection time, except for a small decay 
due to the increasing radiaton field in the source, which increases 
the particle cooling. 
This is shown by the first three distributions in Fig. \ref{fig1}, labelled 
according to the time (in $R/c$ units) after the beginning of the injection.

Fig. \ref{fig2} shows the synchrotron self--Compton spectra 
produced by the distributions in Fig. \ref{fig1},
from $t=0.02$ to $t=10 R/c$ {\it without considering
light travel time effects}.
The first instants (while the injection is active) are shown 
in Fig. \ref{fig2}a, and the decay in Fig. \ref{fig2}b. 
At frequencies
above the synchrotron peak the cooling time is short,
the equilibrium state is reached quickly and the flux is steady during the
injection time, except for the very initial instants, not reproduced in 
Fig. \ref{fig2}a.
Note that since electrons are injected only at very high energies
their emission is first concentrated at high frequencies; 
only after some cooling time they can substantially emit at lower 
frequencies, simulating a new particle injection at lower energies. 
This can be seen in Fig. \ref{fig2}b, where the flux at $\nu<10^{13}$ Hz
increases for $t>t_{inj}$.

Fig. \ref{fig3} shows the spectra produced by considering 
the effect of the {\it light crossing time}: 
as we already emphasized in section \ref{slices} this effect
must be included when the $N(\gamma)$ distribution evolves on time scales 
shorter than $R/c$.
In this case the observer will see different emissions coming from
different regions of the source, {\it as the source itself were not 
homogeneous}.
We stress that this is only due to the photon crossing time.
In order to properly calculate the {\it observed} spectral evolution, 
we have to sum the correct contributions of the different source slices.
The spectra shown in Fig. \ref{fig3} correspond to the first 
instants of the evolution.
Note that the intensity of synchrotron radiation is now increasing during the
injection time also at frequencies above the peak. 
This is mainly due to the photon crossing time: 
when the electron cooling time is very short with respect to $R/c$ 
(as it is for example
for electrons emitting at the synchrotron peak at $\nu\sim 10^{17}$ Hz),
the equilibrium state is reached quickly, and during the
injection the observer receives photons coming from an increasing volume.
After $t=t_{inj}$ the flux remains 
steady until $t=R/c$, as can be easily seen also in the light curves 
(\S \ref{glc}, Fig. \ref{fig4}).
This effect is also shown by the Compton component.

Due to the rapid cooling, electrons at the highest energy 
reach equilibrium in a short time, and during the injection time
the distribution at these energies remains steady.
After $t_{inj}$, it decays very rapidly.
The corresponding synchrotron emission is then `switched on' for a short
time (during which it is constant) and then it is `switched off'.
At any given time (within $R/c$) the observer will then see a constant flux,
produced by a single `switched on' slice 
(which is `running' across the source).
This behaviour can be also seen in Fig. \ref{fig4} where the light curves
at different frequencies are shown.

\subsubsection{\it Light curves}
\label{glc}

\begin{figure}
\psfig{file=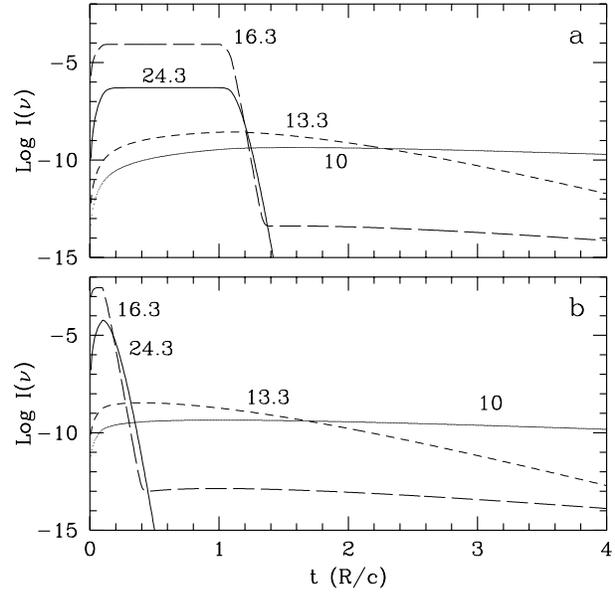,width=8.5truecm,height=8.5truecm}
\caption[h]{Light curves of the specific intensity at different frequencies, 
corresponding to the spectral evolution of Fig. 2, to illustrate
light crossing time effects, included in a) and ignored in b).
Labels correspond to the logarithm of the frequency,
as seen in the comoving frame. For clarity, the 
intensity of each light curve has been multiplied by different constants.}
\label{fig4}
\end{figure}

We can compare the light curves obtained considering or neglecting light
crossing time effects.
This is shown in Fig. \ref{fig4}a and Fig. \ref{fig4}b, respectively.
We choose three frequencies on the synchrotron component as they are 
characteristic of the three different energy ranges, and an additional 
frequency around the (initial) Compton peak. 

Fig. \ref{fig4}a shows that at very low frequencies, 
(i.e. $\nu\sim 10^{10}$ Hz), where $t_{cool} \gg R/c$, the rise
of intensity is slow, because it is mainly controlled by the the
light crossing time $R/c$: the observer will see the first slice, then the
first and the second slices, etc.; when the last slice become visible, the
first one is still emitting, because particle cooling is very slow.
Since electrons are injected mainly at high energies, they can 
substantially emit at lower frequencies only after some cooling time scales:
after $t=R/c$ the intensity is still rising in some slices,
causing the flux at these low frequencies to peak at $t>R/c$.

The intensity at higher frequencies peaks instead at earlier times.
At very high synchrotron frequencies ($\nu\sim 10^{16}$ Hz and above) we will 
have a plateau, because both $t_{cool}$ and $t_{inj}$ are
much shorter than $R/c$: the electrons cool so fast that we will see a single 
slice `running' across the source, as mentioned above.


The shape of the decay phase 
is different for different frequencies: it is mainly controlled
by $t_{esc}$ at very low frequencies (see the light curve at 
$\nu=10^{10}$ Hz), while cooling dominates at higher frequencies.

Note that the decay in the low frequency light curves in Fig. \ref{fig4}a 
are very similar to those in Fig. \ref{fig4}b, while the entire high frequency
light curves and the initial instants of the low frequency ones cannot be
reproduced without including the light crossing time effect. 
This happens because in these last two cases $R/c$ is longer than $t_{cool}$ or
the injection time, respectively.

In Fig. \ref{fig4} we also show the $\nu=2\times 10^{24}$ Hz light curve,
which is close to the (initail) peak of the Compton emission component.
The shape of this curve is very similar to that of the $\nu=2\times 10^{16}$
Hz one, which is close to the synchrotron peak, because emission mainly comes 
from electrons of approximately the same energy.

\subsection{Power--law injection with $t_{inj}=R/c$}
\label{pl}

We now show the case of electrons distributed in energy as 
a power law $Q(\gamma)\propto
\gamma^{-p}$ ($p>0$), continuously injected for $t_{inj}\sim R/c$.
Input parameters for this case are: 
$R=10^{16}$ cm, $Q(\gamma)\propto \gamma^{-1.7}$
cm$^{-3}$ s$^{-1}$ with $\gamma_{min}=1$, $\gamma_{max}=10^5$,
$\ell_{inj}=10^{-3}$, $B=1$ Gauss.

\begin{figure}
\psfig{file=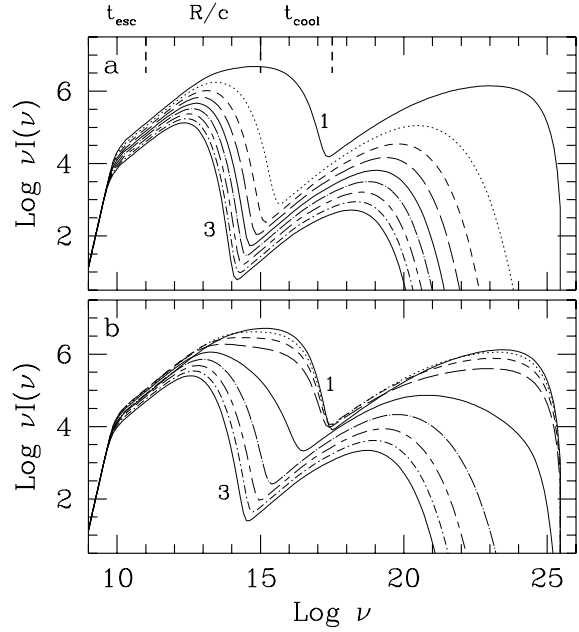,width=8.5truecm,height=8.5truecm}
\caption[h]{Evolution of the synchrotron self-Compton
spectra calculated assuming to inject a power-law distribution,
for $t_{inj}=R/c$.
In a) light crossing time effects have been neglected, while in b)
they have been taken into account. 
For clarity, only the decaying phase is shown.
The shown spectra are separated in time by $\Delta t=0.25R/c$.
Top labels divide the frequency range according to the relevant
time scales (see text).}
\label{fig5}
\end{figure}

We plot the time dependent 
spectra from $t=R/c$ to $t=3 R/c$ in Fig. \ref{fig5}a without considering 
the effect of the slices and in Fig. \ref{fig5}b we summed the correct 
contribution of the different slices.

Also in this case we roughly divide the synchrotron component into 
three energy ranges:
\begin{enumerate}
\item for $\nu<10^{11}$ Hz the corresponding electron cooling time
is much longer than $R/c$, and the decay of the flux (for $t>t_{inj}$) is
mainly due to particle escape;
\item for $\nu>10^{15}$ Hz we have $t_{cool}\ll R/c$: this is the part of
the emitted spectrum in which we can see the greatest differences between the
behaviours reported in Fig. \ref{fig5}a and \ref{fig5}b; in other
words, in this range of frequencies it is strictly necessary to take into
account the different photon crossing times in order to correctly
reproduce the evolution of the emitted spctrum;
\item in the middle region we have $t_{cool}\sim R/c$.
\end{enumerate}

It is interesting to compare the flux behaviour, after the stop of the 
injection, in the spectral region in which $t_{cool}\ll R/c$:
in Fig. \ref{fig5}a (light crossing time effects ignored)
we see that at the highest synchrotron frequencies 
the decay is very rapid, since the entire source closely follows 
the decay of the corresponding electron distribution.

\begin{figure}
\psfig{file=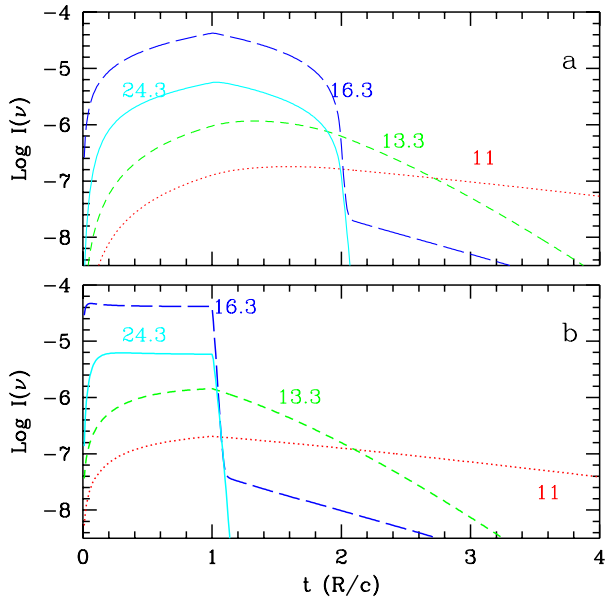,width=8.5truecm,height=8.5truecm}
\caption[h]{Light curves of the specific intensity at different frequencies, 
corresponding to the spectral evolution of Fig. 5, to illustrate
light crossing time effects, included in a) and ignored in b).
Labels correspond to the logarithm of the frequency, as
seen in the comoving frame. 
For clarity, the intensity of each light curve has been multiplied by 
different constants.}
\label{fig6}
\end{figure}

In the second case (Fig. \ref{fig5}b, taking into account light crossing time
effects) the behaviour is more complex, and after the end of the 
injection the emission falls showing three different phases:
\begin{itemize}
\item $t \gta R/c$: we have a slow decay; we receive
the emission from the distribution that starts to cool from the front
slices, while the back slices are still `switched on'.
\item $t\sim 2R/c$: when almost the entire source is (seen) cooled, 
the decay speeds up, because the only important contributions 
are given by the back slices, 
\item $t\sim 3R/c$: the entire source will not contain highly energetic 
electrons any longer, and the emission at $\nu> 10^{15}$ Hz is dominated
by the Compton component, produced by electrons of relatively small energy,
whose cooling time is long.
Correspondingly, at these frequencies the decay is again slow.
\end{itemize}
This behaviour is illustrated by the light curves shown in Fig. \ref{fig6}.
 
For the flux at the highest (Compton) frequencies, the behaviour is similar
to the synchrotron flux, up to $t\sim 2 R/c$, after which the flux disappears.

It is important to note 
that just after the end of the injection, when high energy electrons are 
present and synchrotron photons are scattered at $\nu \ge 10^{24}$ Hz, 
we find similar variability factors for frequencies above the two peaks,
contrary to the `normal' behaviour which is characterized by
a quadratic dependence of the Compton flux with respect to the
synchrotron flux.
This is due to the fact that the Compton scattering at these 
frequencies is in the Klein Nishina regime: high energy electrons
efficiently scatter seed photons of frequencies 
$h\nu/m_ec^2 < 1/\gamma$ (see equation 16).
For the highest energy electrons the available synchrotron photons
for scattering are less, reducing the variability amplitude
(see e.g. Ghisellini, Maraschi \& Dondi 1996).

\subsubsection{\it Light curves}

We report in Fig. \ref{fig6}a and \ref{fig6}b the light curves at four
different frequencies considering or not light crossing time effects.
We can see that taking them into account we can have symmetric light curves,
with equal rising and decaying phases, at frequencies for which 
$t_{cool}\ll R/c$. 
The cooling time of the electrons emitting at high frequencies 
($10^{15}-10^{17}$ Hz) is much shorter than $R/c$, and the electron
distribution remains steady for the entire injection time.
This implies that for $t<R/c$ the observer sees an increase of the
flux due to the increasing `switched on' volume of the source.
After $t=R/c$, the front slices `switch off', and the total flux decreases.
This corresponds to symmetric light curves without plateaux.
At lower frequencies, at which $t_{cool}>R/c$, the decay is slower than 
the rise, and the light curves are asymmetric.

Another remarkable effect is the existence of time-lags among the light 
curves at different frequencies: it is easy to note that the emission
peaks are not reached at the same time and the flux at the highest 
frequencies appear to lead.

All these behaviours are not observable if we do not take into account 
the light crossing time: in this case we see the formation of a plateau 
at high frequencies, where the equilibrium state is reached first. 
Furthermore we cannot observe time lags
between different frequencies, because when the injection stops,
the intensity begins to fall at all frequencies, 
although with different decay shapes (steeper at higher frequencies).

In Fig. \ref{fig6}a we can observe the changes in the 
slope of the decay of the $\nu=10^{16.3}$ Hz light curve, according
to what discussed in Sect. 3.2.


\subsection{Time delays}

We have already noticed the existence of time-lags among light curves at 
different frequencies, that are clearly visible in Fig. \ref{fig6}a: 
this is a consequence of taking into account the light crossing time effects,
although time delays are mainly caused by the different cooling times of 
electrons emitting at different frequencies. 

Assume in fact that $t_{inj}=R/c$:
at $t=t_{inj}$ the observer sees the stop of the particle injection in 
the `front' of the source, while the back of the source is still 
`switched off'; in other words, at this time the observer is receiving the 
spectra emitted by a population of electrons at $t_{evol}=R/c$ 
(from the front of the source) and at $t_{evol}=0$ 
(from the back of the source). 
At later times in the front of the source the electron distribution 
starts to decay (with a corresponding decay of the emitted flux), 
while in the back the flux is still rising until the particle distribution 
reaches, after some $t_{cool}$, the equilibrium state.
This combination of slices in which the emitted flux is increasing and 
slices in which the decaying phase is already started, determines the 
position of the peak on each light curve. 
Notice that if the electron cooling time is very short 
(for example in the case of the $\nu=10^{16}$ Hz in Fig. \ref{fig6}), 
the equilibrium state is reached in a very short time with respect to $R/c$: 
after the injection stops, each slice is quickly {\it turned off}, 
and the peak of the light curve is reached soon after $t=t_{inj}=R/c$.
In light curves corresponding to lower energy electrons the peak will 
occour later, because the equilibrium state is reached later:
after the stop of the injection the flux is slowly decreasing in the front 
of the source, while it is still increasing in the back. 
The longer the particle cooling time, the longer the time that
the light curve needs to reach the maximum.

We emphasize that the different cooling times of the electrons
responsible for the emission at different frequencies introduce 
more visible time--lags if the light crossing time effects are included,
as can be seen comparing the light curves for the high synchrotron 
frequencies shown in Fig. 6a and Fig. 6b.

\subsection{Simulation of a shock}
\label{shock}

We assume that a shock of longitudinal dimension $R$ and width $r \ll R$
runs along a region of the jet of same dimension $R$ (perpendicular to the 
jet axis). 
Particle flows along the jet with a velocity $\beta c$.
In the frame comoving with this flow the shock moves with a velocity 
$\beta^{\prime}_s c$. 
As in the previous sections, we assume that the observer is located 
at an angle $1/\Gamma=\sqrt{1-\beta^2}$ from the jet axis, such that 
the viewing angle in the comoving frame is $90^\circ$. 
We assume that the shock is active for a time $t_s$, as measured
in the comoving frame.
At a given location, the shock accelerates particles for a time 
$t_{inj}=r/(\beta^\prime_s c)$.

\begin{figure}
\psfig{file=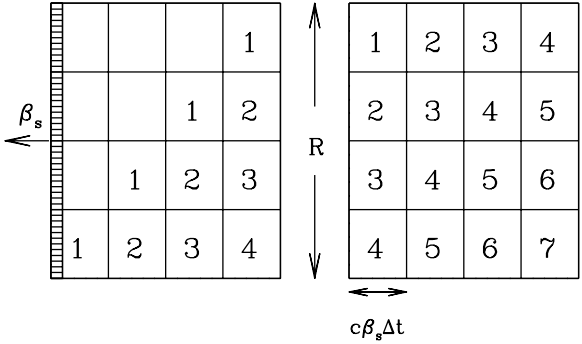,width=10truecm,height=10truecm}
\vskip -4 true cm
\caption[h]{Schematic illustration of the composition of
spectra arriving at the observer at the same time, but leaving
each shell at a different time, hence produced by electrons
of a different age.
The observer is assumed to be located at $90^\circ$ to the shock
velocity (bottom of the figure) in the frame comoving with the 
particle flow.
In the left panel the shock has been active for a time $R/\beta^{\prime}_s c$
since the beginning of the injection.
The number in each cell indicates the ``age" of the electrons.
Note that above the ``diagonal" line (cells without number), 
photons had not enough time to reach the observer.
The right panel shows how to sum the contributions of the cells
at the time equal to $2R/\beta_s^{\prime} c$, in the case that 
the shocks stops to inject particles after $t_s =R/c$.
At later times each cell will contribute to the emission
with an older distribution, and correspondingly the number
characterizing each shell will increase by one unit at each
timestep $\Delta t$.}
\label{jet}
\end{figure}

\begin{figure}
\psfig{file=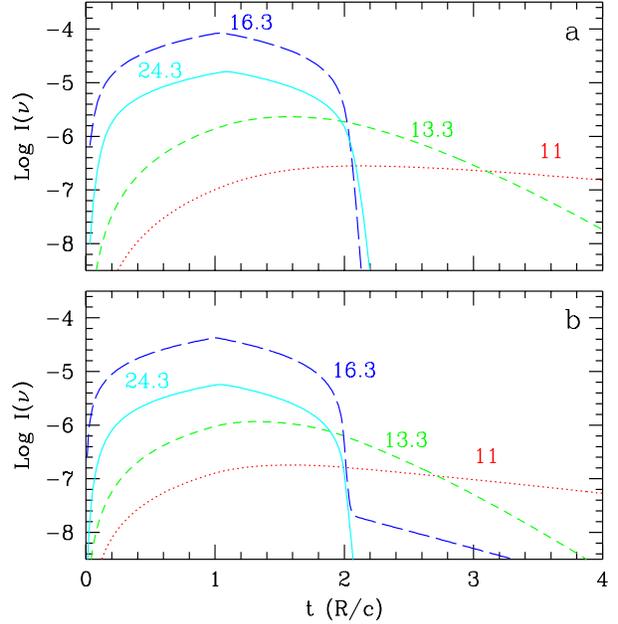,width=8.5truecm,height=9truecm}
\caption[h]{a) Light curves at different frequencies (as seen
in the comoving frame) corresponding
to a simulation of a shock of width $r= t_{inj} \beta^{\prime}_s c$,
$t_{inj}= 0.1 R/c$, injecting energetic particles
for a time $t_{s}=R/c$. 
All other parameters are the same as in the simulations
shown in Fig. 6. In panel b) we report the the light curves already shown
in Fig. 6a in the case $t_{inj}= R/c$. For clarity, the intensity of each 
light curve has been multiplied by different constants.}
\label{sh}
\end{figure}

To show how to correctly sum the different contributions of different emitting
regions, we sketch in Fig. \ref{jet} an illustrative example with a small
number of different regions. In this case, each slice parallel to the jet axis
is characterized by different electron distributions: `young' electrons close
to the shock front, and older ones far from it. It is therefore necessary to 
subdivide each slice into an appropriate number of `cells' (see Fig.
\ref{jet}).
The velocity of the shock in the frame comoving with the particle flow 
determines the size of the cells in the direction parallel to the jet axis,
and then the total emitting volume. 

Fig. \ref{sh}a shows the light curves obtained in the case of
a shock of width $r = t_{inj}\beta^{\prime}_s c$, with $t_{inj}=0.1 R/c$ 
active for a time $t_{s}=R/c$, with all other parameters equal to the case 
of \S \ref{pl}.

Comparing Fig. 8a with Fig. 8b one can notice that in the shock case we obtain 
further delays between different frequencies, due to the different
way of summing the spectra produced at different timesteps. 

From these simulations we can conclude that there are no qualitative 
differencies between the light curves in the `homogeneous' case with 
$t_{inj} =R/c$, except for the occurrence of further time 
delays between light curves at different frequencies.

\subsection{Multiple rapid injections}

\begin{figure}
\psfig{file=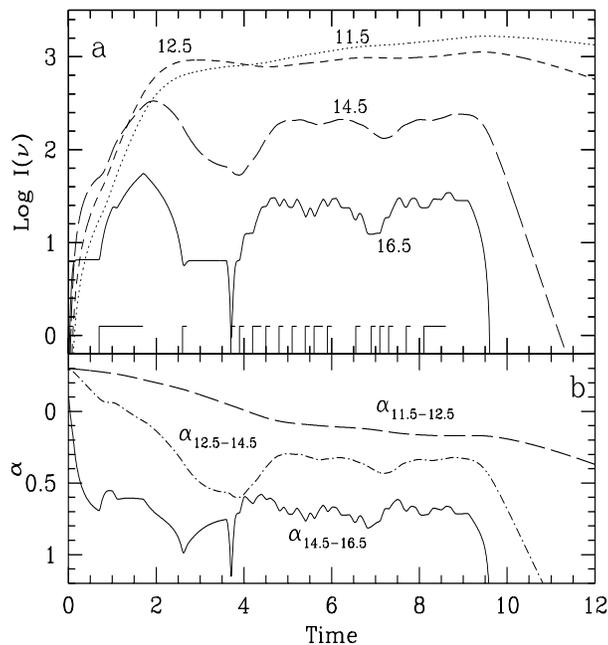,width=8.5truecm,height=10truecm}
\caption[h]{Light curves of the specific intensity
at different frequencies (as seen in the comoving frame) and spectral 
indices resulting from multiple injection phases. For clarity, the 
intensity of each light curve has been multiplied by different constants.}
\label{fig7}
\end{figure}

In Fig. \ref{fig7}a we show the light curves resulting from
discontinuously injecting a gaussian distribution of particles,
centered at $\gamma=10^5$.
In the shown case, we have assumed that there are 17 injection phases,
lasting from $t_{inj}=0.1 R/c$ to $t_{inj}=R/c$, separated by 
different times, and that $t_{esc}=8 R/c$.
The electron injection rates are equal for all cases.

It can be seen that these injection phases can be resolved at the highest 
synchrotron frequencies, while they merge at lower frequencies, where
the memory of the different injection phases is lost.
At energies at which $t_{cool}>R/c$, the electron distribution builds up, 
since electrons cool in a time scale longer than the time separation
between different injection phases.
This introduces both a smoothing of the corresponding light curve
and a time delay with respect to higher frequency light curves.

The series of short injection phases separated by short times
makes the synchrotron flux at the highest frequencies to flicker.
The superposition of the different distributions corresponding to each
injection phase prevents the formation of the `plateau', characteristic
of a single short time injection.
This also causes the variability time scales to be shorter than $R/c$.

In Fig. \ref{fig7}b we show the time evolution of 
three spectral indices 
($\alpha_{11.5-12.5}$, $\alpha_{12.5-14.5}$ and $\alpha_{14.5-16.5}$, 
connecting the flux 
at $\log \nu=11.5$ and $\log \nu=12.5$, the flux
at $\log \nu=12.5$ and $\log \nu=14.5$ and the flux
at $\log \nu=14.5$ and $\log \nu=16$, respectively, where
frequencies are measured in Hz).
The spectral indices can vary (even by a large amount) during
the major injection phases, while they remain within a narrow range
during the flickering phases.

\section{Discussion}

We summarize here the main characteristic of the light curves for
different values of $t_{inj}$, $t_{cool}$ and $t_{esc}$:
\begin{itemize}
\item $ t_{inj}\ll R/c $
\begin{itemize}
\item when $t_{cool}\gg R/c $ we see a slow rise of intensity,
mainly controlled by the light crossing time $R/c$: the maximum of intensity
is reached for $t>R/c$, and depends on $t_{cool}$ and $t_{esc}$.
At small frequencies, where $t_{cool}\gg t_{esc}$, the decay of the emission
is controlled by $t_{esc}$, while at frequencies where 
$t_{cool} \le t_{esc}$ the leading effect is cooling;

\item  when $t_{cool} \simeq R/c$ the rising shape of the light curve
is similar to the above case.
The maximum of intensity is now reached at $t\sim R/c$,
and the shape of the decay is controlled by $t_{cool}$; in the special 
case in which $R/c\sim$ some $t_{cool}$ we have a symmetric light curve;

\item if $t_{cool}\sim t_{inj}$ we have
a fast rise of intensity ($t_{rise}\simeq t_{inj}$) and then a plateau (or a 
very slow rise in the shock case) until $t\sim R/c$. 
For $t>R/c$ the decay is as fast as the initial rise.

\end{itemize}
\item $t_{inj}\sim R/c$
\begin{itemize}
\item if $t_{cool} \gg R/c$ we see a slow rise, the maximum is 
at $t\approx 1.5 R/c$, and the decay is due to escape if $t_{cool}\gg t_{esc}$;
\item if $t_{cool}\sim t_{inj} \sim R/c$ the shapes of the rising and decaying
phases are very similar, resulting in a symmetric light curve;
\item where $t_{cool} \ll R/c$ both the rise and the decay are controlled
by the light crossing time, and we have symmetric light curves with
time lags depending on the different cooling times;
\end{itemize}
\item if $t_{inj}$ is much longer than $R/c$ and $t_{cool}$ the light curve
will have a plateau, because the long injection time allows the entire
source to reach equilibrium.
\end{itemize}

The main difference between injecting electrons at high energies
(`monoenergetic' or gaussian injection) and injecting a power--law 
(with $p>0$) distribution is that in the first case the emission will be 
concentrated first at high frequencies, and only after some $t_{cool}$ 
electrons can substantially emit at lower frequencies.
This simulates a sort of a `new' injection at low energies, that
continues for $t>t_{inj}$, and produces a time delay between the peak
of the emission at different frequencies.

If the emitting plasma has bulk motion, we must take into 
account the effects of beaming.
If $\Gamma$ is the bulk Lorentz factor, $\theta$ the viewing angle and
$\delta=[\Gamma(1-\beta \cos\theta)]^{-1}$ the beaming factor,
the observed intensity is $I(\nu)=\delta^3 I^\prime (\nu^\prime)$
and $t=t^\prime/\delta$,
where $I^\prime (\nu^\prime)$ and $t^\prime$ are the comoving intensity
and comoving time scales, respectively.

Observations of variability in BL Lac objects suggest the simultaneous 
presence of different variability time scales on each source (e.g. Ghisellini
et al. 1997, Massaro et al. 1996), suggesting that the variations
 can be originated
by more than one component: a larger one can take into account of the longer
time scale variations, and smaller ones can originate the rapid flares.
Flares are observed at almost all wavelengths (depending on each source 
characteristics), and can have different durations (from hours to days).
The occurrence of outbursts with symmetric shape (similar rise and fall)
have been recently reported for well monitored sources, especially in the 
optical band (Massaro et al. 1996, Ghisellini et al. 1997) and in the 
X--ray band (Urry et al. 1997). 
According to our results, this behaviour can be originated only in two cases:
\begin{enumerate}
\item $t_{inj}\ll R/c \sim t_{cool}$
\item $t_{inj}\sim R/c$ and $t_{cool}\ll R/c$.
\end{enumerate}
In case (i) symmetric light curves are present only within a very small
range of frequencies ($t_{cool}\sim R/c$), while in case (ii) quasi--symmetric 
light curves can occur at all frequencies corresponding to particle cooling 
time scales shorter than $R/c$. 
Furthermore the second case can be interpreted 
as a result of a shock lasting for a time $t_{inj}\sim R/c$.

Note that the underlying component has not been included in the 
previous simulations (\S\S \ref{gauss} and \ref{pl}), which only 
describe the evolution of the `flaring' component.
To correctly reproduce the blazars variability behaviour, 
it is therefore necessary to take into account the presence of
a quasi-stationary component, {\it diluting} the variability of the
flaring component.

Finally, note that the simulations in \S 3.4 of a shock active for
a time $R/c$ are similar to the corresponding `homogenous' case,
with slightly longer time delays between light curves at
different frequencies.

\section{Application to MKN 421}

Mkn 421 is one of the nearest BL Lac Lac objects ($z=0.03$)
and it is classified as 
XBL (X--ray selected BL Lac) (Giommi et al. 1990, Hewitt \& Burbidge 1993). 
In the overall spectral energy distribution [$\nu-\nu F(\nu)$] the 
X--ray emission smoothly connects to the optical and UV. 
It is generally believed that this component is due to synchrotron emission. 
Mkn 421 is a faint EGRET source with a flat GeV spectral index 
(Sreekumar et al. 1996) and it was the first blazar detected 
at TeV energies (Punch et al. 1992). 
In the optical, UV, X--ray and TeV bands it shows strong and rapid 
flux variability (Buckley et al. 1996).

\begin{figure}
\psfig{file=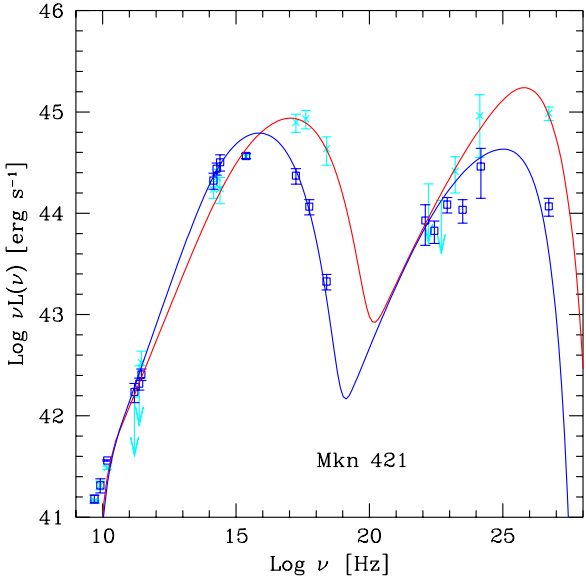,width=8.5truecm,height=8.5truecm}
\caption[h]{High and low states of Mkn 421 during the 1994 multiwavelength
campaign (Macomb et al. 1995); continuous lines are equilibrium
SSC models. We found satisfactory fits with the following parameters:
$R=4.8 \times 10^{16}$ cm, $B=0.08$ Gauss,
injection law $Q(\gamma)\propto \gamma^{-1.5} \exp(-\gamma/\gamma_{max})$
with $\gamma_{min}= 1$, 
$\gamma_{max}=1.3\times 10^5$, $t_{esc}=4 R/c$, injected compactness
$\ell_{inj} = 9 \times 10^{-5}$, beaming factor $\delta=15.5$
{\it (low state)};
$R=4.8 \times 10^{16}$ cm, $B=0.04$ Gauss, $Q(\gamma)\propto \gamma^{-1.7}
\exp(-\gamma/\gamma_{max})$ with $\gamma_{min}= 1$, $\gamma_{max}=8.5\times 
10^5$, $\ell_{inj}=2.5\times 10^{-4}$,
$t_{esc}=3 R/c$, $\delta=15.5$ {\it (high state)}.}
\label{mkn_hl}
\end{figure}

In May 1994 the ASCA satellite revealed an X--ray flare  
(Takahashi et al. 1996) during an high state of TeV emission 
(Macomb et al. 1995). 
Observations report an increase of a factor $\sim 2$ of the 2--10 keV flux,
with a doubling time scale of $\sim 12$ hours, while much less amplitude
variability is present in the IR, optical, UV and GeV (EGRET) bands. 
Takahashi et al. (1996) found a time-lag between hard X--rays 
($2-7.5$ keV) and soft X--rays ($0.5-1$ keV) of $\sim 1$ hour: the hard 
X--rays lead the soft X--rays.
They interpret this as due to synchrotron cooling.
With the numerical code described in the previous sections we can qualitatively
reproduce this behaviour, first by fitting the high and the low states (Macomb
et al. 1995) as equilibrium SSC spectra, and then by generating 
simulated light curves at all frequencies. 
In Fig. \ref{mkn_hl} we report the simultaneous spectra taken during 
the high and low states (data from Macomb et al. 1995, 1996, using $H_0=50$ Km 
s$^{-1}$ Mpc$^{-1}$, $q_0=0.5$). 
The parameters used for the fits are reported in the figure caption. 

We assumed that the rapid variability revealed by ASCA during the highly
active X--ray/TeV state is due to the sum of a rapidly evolving component 
and a quasi-constant one, corresponding to the high state fit. 
We add to this constant emission a variable component, for which we found
the following parameters:

$R=1.5\times 10^{16}$ cm, $B=0.13$ Gauss, $\delta=15.5$, 
$\ell_{inj}=1.5 \times 10^{-3}$, $Q(\gamma)\propto \gamma^{1.4} 
\exp(-\gamma/\gamma_{max})$ between 
$\gamma_{min}=10^3$ and $\gamma_{max}=8.5\times 10^5$. We perform the
simulation in the shock case with the following parameters:
$r_s= 0.1 R/c$, $t_{s}= R/c$ and $\beta^{\prime}_s \sim 1$.

\begin{figure}
\psfig{file=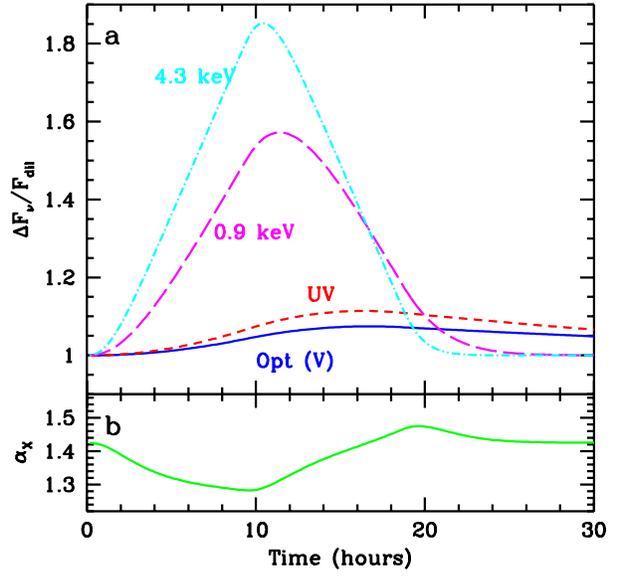,width=8.5truecm,height=8.5truecm}
\caption[h]{Ligh curves at different energies, reproducing the rapid 
flare shown by MKN 421 on May 16, 1994. In a) we plotted
the flux at different frequencies normalized to the steady component; in b)
the spectral index connecting the 4.3 keV and the 0.9 keV fluxes is shown. 
Times and frequencies take into account of the beaming factor $\delta$.}
\label{mkn_lc}
\end{figure}

We plot the obtained light curves at four different frequencies 
in Fig. \ref{mkn_lc}. 
Note that the flux is still increasing at all reported frequencies for
$t > t_{s} (1+z)/\delta$, which corresponds to $t\sim 9$ hours: this is 
due to the different evolving state of the electrons in the different cells.

Because of the photons travel time, when the injection stops (at $t_s =R/c$) 
the observer still sees the emission from the back of the source
produced by the $increasing$ particle distribution.
Therefore the maximum flux is observed at a time greater than $R/c$,
by an amount depending on the cooling time: the flux at the highest frequencies
peaks first, followed by the flux at lower frequencies.
As a consequence, the flux at the synchrotron peak (X--ray band) 
leads the flux at lower frequencies, and time-lags can be observed even 
between close frequencies within the same band. 

Another consequence of our modelling is the prediction of time-lags 
between the synchrotron and the self-Compton fluxes.
While the X--ray and the TeV emission are probably produced by the same
electrons, the most effective seed photons for the TeV emission are in 
the (observed) optical--UV band, due to the Klein Nishina decline of 
the scattering cross section.
In the case discussed here, the electron population producing the 
optical--UV emission builds up, as a consequence of particle cooling,
some time after the beginning of the injection.
Optical--UV and TeV emissions should therefore be nearly simultaneous,
lagging the X--ray emission.
A delay between X--ray and TeV emission can also be introduced by taking 
into account the different light travel times of the seed photons to be 
scattered at high energies, as mentioned in \S 2.3, but to be more
quantitative we must await a more detailed numerical treatment.

A time lag between X--ray and TeV fluxes should exist
only if the seed photons for the scattering are produced locally.
In the alternative case of seed photons coming from external regions
(by, e.g. a dusty IR torus), we expect to observe a simultaneous
rise in the X--ray and TeV fluxes of the same factor.
Note that the `external' radiation could also come from adiacent
regions of the flaring component, i.e. the ones responsible for
the `diluting' component.
Also in this case the X--ray and the TeV fluxes, being produced
by the same electrons, should vary linearly and simultaneously.

\section{Conclusions}

By summing the contribution of different slices of the source, the
observer receives photons produced by particle distributions
of different ages, a situation which resembles the one occuring
in an inhomogeneous source.
This effect is important whenever the particle distribution evolves
on time scales shorter than $R/c$.
Despite the fact that this effect tends to smooth out fast variations,
time delays between light curves at different frequencies are observable,
as illustrated for the hard/soft X--ray emission of Mkn 421.
Our results are particularly important for the observed fast variability of 
blazars, where the variability time scales indicate extremely compact 
emission regions, thus large magnetic and radiation energy densities, and 
consequently cooling time scales shorter than the light crossing time $R/c$.
This is also a necessary condition for having symmetrical flares 
(equal rising and decay time scales), often observed in the optical and in 
the X--ray band. 
Note that the case of 3C 279, one of the best studied blazars which 
showed asymmetric light curves at high energies, is probably different,
because the high energy emission can be inverse Compton scattering
off seed photons produced externally to the jet, and the resulting
light curve can be connected to the geometrical distribution
of these seed photons: for example, if an active blob passes
through the broad line region (BLR), there will be enhanced inverse
Compton emission as long as the blob is inside the BLR, and a sudden
drop when the blob moves outside the BLR.

Within the synchrotron self--Compton scenario,
the symmetrical behaviour of the light curves
near or above the peak of the synchrotron emission  
can be contrasted with the light curves at lower frequencies,
characterized by cooling (and escape) time scales longer than $R/c$.
In this case the light curves should be asymmetric, with a rising
phase lasting $\gta R/c$ (if $t_{inj}\sim R/c$), and a decaying
phase lasting for $t_{cool}$ (or $t_{esc}$, if shorter than $t_{cool}$).
Sources with different magnetic field and/or different compactnesses should 
behave differently, since for large compactness (or larger magnetic field) 
sources, the cooling time can be shorter than $R/c$ for a wider range of 
electron energies, resulting in symmetric light curves for a wider range of 
frequencies.

We stress that symmetric light curves without plateaux strongly constrain
the injection and the cooling time scales:
if both these times are much longer than the the light crossing time
$R/c$ the light curve is not symmetric;
if both time scales are much shorter than $R/c$ the curve is symmetric, 
but a plateau forms,
while if the injection time is shorter than $R/c$ and the cooling time
is of the same order, a symmetric light curve is possible (without plateau),
but only at one specific frequency.
Therefore symmetric light curves without plateau 
at more than one frequency are possible only if the the injection
last for $R/c$ and the cooling time is shorter than $R/c$.

A more complex behaviour is possible if the injection of relativistic
electrons is impulsive and repeated several times within one
light crossing time $R/c$.
In this case the synchrotron flux at the largest frequencies responds
to the different injection phases, and the repeated injections can
make the plateaux in their light curve to disappear.
Variations of relatively small amplitude are possible in very short times, 
which would lead to calculate variability time scales shorter than $R/c$.
At lower frequencies, where cooling times are longer, the electron 
distributions corresponding to different injections can build up
and the memory of the individual injection phases is lost.
At these frequencies the light curves are smoother, and they
can have apparent delays (with respect to higher frequency light curves)
produced by this `build up' effect.

\section*{Acknowledgments}
We thank G. De Francesco for help during the preparation
of the numerical code, G. Bodo, A. Celotti and L. Maraschi
for useful discussions.

\section*{References}

\refitem Atoyan, A.M. \& Aharonian, F.A., 1997, in Relativistic Jets 
in AGNs, eds. M. Ostroski, M. Sikora, G.M. Madejski \& M. C. Begelman
(Cracow), p. 324

\refitem Bucley, J.H. et al., 1996, ApJ, 472, L9

\refitem Chang, J.S. \& Cooper, G., 1970, Journal of Computational Physics 6, 1

\refitem Crusius, A. \& Schlickeiser, R., 1986, ApJ, 164, L16 

\refitem Ghisellini, G., Guilbert, P. \& Svensson, R., 1988, ApJ, 334, L5 

\refitem Ghisellini, G. et al., 1997, A\&A, 327, 61

\refitem Ghisellini, G. \& Svensson, R., 1991, MNRAS, 252, 313

\refitem Ghisellini, G., Maraschi, L. \& Dondi, L., 1996, ApJS, 120, 503

\refitem Giommi, P. et al., 1990, ApJ, 356, 432 

\refitem Giommi, P. et al., 1998, A\&A, 335, L5

\refitem Hewitt, A. \& Burbidge, G., 1993, ApJS 87, 451

\refitem Kardashev, N.S., 1962, Soviet Astronomy-AJ, 6, 317

\refitem Macomb, D.J. et al., 1995, ApJ, 449, L99

\refitem Macomb, D.J. et al., 1996, ApJ, 459, L111 (Erratum)

\refitem Maraschi, L., Ghisellini, G. \& Celotti, A., 1992, ApJ, 397, L5

\refitem Massaro, E. et al., 1996, A\&A, 314, 87

\refitem Mastichiadis, A. \& Kirk, J.G., 1997, A\&A, 320, 19

\refitem Press, W.H. et al., 1989, Numerical Recipes in Fortran, Cambridge
University Press

\refitem Punch, M. et al., 1992, Nature, 358, 477. 

\refitem Rybicki, G. \& Lightman, A.P., 1979, Radiative Processes 
in Astrophysics, Wiley Interscience, New York

\refitem Sreekumar, P. et al., 1996, ApJ 464, 628

\refitem Takahashi, T. et al., 1996, ApJ 470, L89

\refitem Ulrich, M.-H., Urry, C.M. \& Maraschi, L., 1997, ARAA, 35, 445

\refitem Urry, C.M. et al., 1997, ApJ, 486, 799

\refitem Zdziarsky, A.A., 1986, ApJ 305, 45

\end{document}